\documentclass[preprint]{ptephy_v1}

\usepackage{booktabs}
\usepackage{multirow}
\usepackage{color}
\usepackage{graphicx}
\usepackage{dcolumn}
\newcolumntype{d}[1]{D{.}{.}{#1}}
\usepackage{bm}
\usepackage{braket}
\usepackage{amsmath}
\usepackage{here}
\usepackage{color}
\usepackage{array}
\usepackage{enumerate}
\usepackage{amssymb}
\usepackage{bbm}
\usepackage{dsfont}
\usepackage[scr=rsfs]{mathalpha}

\def\be{\begin{equation}} 
\def\ee{\end{equation}}

\usepackage{xcolor}
\graphicspath{{figs/}}
\usepackage[bookmarks=true,colorlinks,%
            citecolor=blue,linkcolor=blue,anchorcolor=blue,filecolor=blue,urlcolor=blue,%
           ]{hyperref}

\usepackage[normalem]{ulem}

\renewcommand\sout{\bgroup\color[rgb]{1,0.75,0.8} \ULdepth=-.5ex \ULset}

\begin{document}

\title{Generator coordinate method with proton--neutron pairing fluctuations
and magnetic properties of \texorpdfstring{$N=Z$}{N=Z} odd--odd nuclei}

\author[1]{K. Uzawa}
\author[2,3,4]{N. Hinohara}
\author[2,3,5]{T. Nakatsukasa}
\affil[1]{
Department of Physics, Kyoto University, Kyoto 606-8502, Japan}

\affil[2]{
 Center for Computational Sciences, University of Tsukuba, Tsukuba, Ibaraki 305-8577, Japan
}
\affil[3]{
 Faculty of Pure and Applied Sciences, University of Tsukuba, Tsukuba, Ibaraki 305-8571, Japan
}
\affil[4]{Facility for Rare Isotope Beams, 
              Michigan State University, East Lansing, Michigan 48824, USA}

\affil[5]{
RIKEN Nishina Center, Wako 351-0198, Japan
}

\date{\today}

\begin{abstract} 
Pairing correlations play an important role in a variety of nuclear phenomena.
However, a quantitative understanding of proton--neutron $(pn)$ pairing, especially isoscalar $pn$ pairing $(S=1, T=0)$ remains elusive.
To clarify the property of $pn$ pairing, we investigate the roles of $pn$ pairing in the $M1$ transition of $N=Z$ odd--odd nuclei.
We develop a theoretical model based on the generator coordinate method (GCM) in which
the isoscalar and isovector $pn$-pair amplitudes are used as the generator coordinates.
Using the particle and the angular-momentum projections,
the $pn$-pair GCM well reproduces the $M1$ transition of odd--odd nuclei
for the exactly solvable SO(8) model.
We apply the method to $N=Z$ odd--odd nuclei and find that the experimental values of $B(M1)$ are well reproduced.
We also study the sensitivity of $B(M1)$ to the strength of the isoscalar pairing interaction.
\end{abstract}

\maketitle

\section{INTRODUCTION}

The pair correlation has significant impacts on a variety of nuclear properties, such as odd--even mass difference, rotational moments of inertia, fission dynamics, and so on \cite{superfluidity}.
The pairing in particles of the same kind (like-particle pairing),
namely, correlations between two protons or between two neutrons that couple to spin singlet $(S=0, T=1)$,
has been extensively studied.
However, protons and neutrons can be regarded as different isospin components of
the same kind of particles ``nucleons.''
It is natural to expect a Cooper pair composed of a proton and a neutron,
leading to the $pn$-pair condensation~\cite{Frauendorf2014}. 

The like-particle pairing is a part of the isovector (IV) pairing with 
($T=1, T_z=\pm 1$).
In the isospin-triplet pairing, there exists a $(T=1, T_z=0)$ $pn$ channel.
The isospin symmetry of the nuclear force implies that
the IV $pn$ pairing may play an important role in $N=Z$ nuclei.
In fact, experimental data 
of the binding energy of $N=Z$ odd--odd nuclei in the $0_1^+$ states
suggest the occurrence of the $T=1$ pair condensation
\cite{Macchiavelli2000}.

On the other hand, our knowledge of the isoscalar (IS) pairing remains limited.
Despite the strong attractive interaction in the $(S=1, T=0)$ channel,
there is no clear experimental evidence for the IS pair condensation.
Nevertheless,
some proton-rich nuclei are suggested to be close to the critical point
\cite{Yoshida2014},
in which the quantum fluctuation associated with the IS-pair vibrations \cite{superfluidity} plays an important role.
Low-energy collective modes have been observed in the Gamow--Teller (GT) energy spectrum in $N=Z$ odd--odd nuclei \cite{Fujita2014,Fujita2019}. 
Theoretical calculations suggest that the low-energy GT peaks develop
as the IS pairing strength increases \cite{Bai2014}.
The deuteron-knockout reaction is also known to provide useful insights into the IS pair correlation.
In Ref.~\cite{Chazono2021} the triple differential cross section of the proton-induced deuteron-knockout reaction $^{16}$O$(p,pd)^{14}$N$^*$ and its sensitivity to the IS pairing strength are studied based on the nuclear density functional theory and the distorted-wave impulse approximation.

Since the IS $pn$ Cooper pair has a nonzero spin ($S=1$),
we expect that the IS pairing correlation influences nuclear magnetic properties,
such as the nuclear magnetic moment and the magnetic transition,
which has been studied theoretically \cite{Tanimura2014,Sagawa2016,Yoshida2021,Jokiniemi2023}.
In this paper, we analyze the role of the IS pairing on the nuclear magnetic properties 
based on a generator coordinate method (GCM) \cite{ring} with both the IV and the IS $pn$-pair amplitudes as the generator coordinates.
We call this method as ``$pn$-pair GCM.''
The collective wave functions obtained in the $pn$-pair GCM
help us to visualize the quantum $pn$-pairing fluctuations.
In order to remove undesirable mixing in the mean-field states,
we combine the GCM with
the projection technique on good total angular momentum ($J$) and
good particle numbers of protons ($Z$) and neutrons ($N$).
In this paper, we apply the method to $N=Z$ odd--odd nuclei. 

If the ground states of proton-rich $N=Z$ nuclei are close to
the critical point of the IS pair condensation,
large-amplitude fluctuation beyond the mean field is important.
In addition, the small-amplitude approximation fails, as
the quasiparticle random-phase approximation (QRPA) collapses
at the critical point.
The GCM adopted in the present study is suitable for the treatment of
such large-amplitude collective motion
\cite{Engel1997,Hinohara2014}.

Among the magnetic properties, we especially focus our study
on the $M1$ transition. 
Although the importance of $(S=1,T=0)$ interaction in the $M1$ transition of $N=Z$ odd--odd nuclei has been discussed based on the three-body model \cite{Tanimura2014},
the relation between the $M1$ transition and IS-pair condensation
has not been perfectly clarified,
due to the limitation of the three-body model.
The $pn$-pair GCM with the projection is able to provide
an intuitive and quantitative answer to the question.
Moreover, the $M1$ transition is dominated by the GT operator of the IV-spin type $(\sigma \tau_z)$, and the present analysis on the spin-flip excitation may lead to the origin of the collective behavior observed in the GT transition \cite{Fujita2014,Fujita2019}.

This paper is organized as follows. 
In Sec.~\ref{sec:HamGCM}, we introduce a pairing model Hamiltonian and
describe the $pn$-pair GCM based on the projected $pn$-mixed mean-field states.
In Sec.~\ref{sec:M1}, we give a brief description of the calculation of $B(M1)$.
Section~\ref{sec:results} provides numerical results.
First, we use the SO(8) (single-shell) Hamiltonian
and compare the result of the $pn$-pair GCM with that of the exact solutions.
Then, we apply the method to the $sd$-shell nuclei
and analyze how the $pn$ pairing affects its nuclear structure and magnetic property.
In Sec.~\ref{sec:summary}, the summary and the conclusion are given.

\section{\texorpdfstring{$pn$}{pn}-pair GCM \label{sec:HamGCM}}
\subsection{Multishell \texorpdfstring{$L=0$}{L=0} pair-coupling model}
In order to treat the IV pairing and the IS pairing on equal footing, we use the following Hamiltonian,
\begin{equation}
\hat{H}=\hat{h}_{\rm spe}
	-G_{\rm IV}\sum_\nu \left(
	\hat{R}^\dagger_\nu \hat{R}_\nu+\hat{R}_\nu \hat{R}^\dagger_\nu
	\right)
-G_{\rm IS}\sum_\mu \left(
	\hat{P}^\dagger_\mu \hat{P}_\mu+\hat{P}_\mu \hat{P}^\dagger_\mu
	\right)
	+G_{\rm ph}\sum_{\mu\nu} \hat{F}^{\mu\dagger}_\nu \hat{F}^\mu_\nu,
	\label{Hamiltonian}
\end{equation}
where $\hat{h}_{\rm spe}$ represents the spherical single-particle energies,
$G_{\rm IV}$ is the IV pairing strength,
$G_{\rm IS}$ is the IS pairing strength,
and $G_{\rm ph}$ is the GT interaction strength.
The IV pair operators $\hat{R}^\dagger_{\nu}$,
the IS ones $\hat{P}^\dagger_{\mu}$,
and GT operators $\hat{F}^{\mu}_\nu$ are defined as follows.
\begin{equation}
	\hat{R}^\dagger_\nu=\frac{1}{\sqrt{2}}\sum_{\alpha,l}\sqrt{2l+1}\left[c^\dagger_{\alpha l}c^\dagger_{\alpha l}\right]_{0 0 \nu}^{0 0 1},
	\quad\quad
	\hat{P}^\dagger_\mu=\frac{1}{\sqrt{2}}
	\sum_{\alpha,l}\sqrt{2l+1}\left[c^\dagger_{\alpha l}c^\dagger_{\alpha l}\right]_{0 \mu 0}^{0 1 0},
\end{equation}
\begin{equation}
	\hat{F}^{\mu}_\nu=\frac{1}{2}
	\sum_{k,k'}
	\bra{k}\sigma_\mu\tau_\nu\ket{k'} c_{k}^\dagger c_{k'}
	=\sum_{\alpha,l}\sqrt{2l+1}\left[c^\dagger_{\alpha l}\bar{c}_{\alpha l}\right]_{0 \mu \nu}^{0 1 1} .
\end{equation}
Here, the single-particle states $c_k^\dag$
are labeled by the LS coupling indices
with $\alpha$ denoting the radial quantum number,
$(l,m)$ denoting the orbital angular momentum,
and $(s,t)$ indexing
the spin $s=\pm 1/2$ and isospin $t=\pm 1/2$.
The index $k$ stands for $(\alpha,l,m,s,t)$.
The $L=0$ coupling products are defined as
\begin{align}
\left[c^\dagger_{\alpha l}c^\dagger_{\alpha l}\right]_{0 \mu \nu}^{0 S T}
	&\equiv \sum_{m,s,t} (l m l -m | 0 0 )
	\left(\frac{1}{2} s \frac{1}{2} \mu-s \Big| S \mu \right)
	\left(\frac{1}{2} t \frac{1}{2} \nu-t \Big| T \nu \right) 
	c^\dagger_{\alpha lmst}c^\dagger_{\alpha l-m\ \mu-s\ \nu-t},\\
\left[c^\dagger_{\alpha l}\bar{c}_{\alpha l}\right]_{0 \mu \nu}^{0 S T}
	&\equiv \sum_{m;s,t} (l m l -m | 0 0 )
	\left(\frac{1}{2} s \frac{1}{2} \mu-s \Big| S \mu \right)
	\left(\frac{1}{2} t \frac{1}{2} \nu-t \Big| T \nu \right) 
	c^\dagger_{\alpha lmst}\bar{c}_{\alpha l-m\ s-\mu\ t-\nu},
\end{align}
where $\bar{c}_{\alpha lmst}=(-1)^{l+1+m+s+t}c_{\alpha l-m-s-t}$.
$\hat{R}^\dagger_1$ and $\hat{R}^\dagger_{-1}$ are commonly used as 
the pair creation operators of neutrons and protons, respectively.
$\hat{R}^\dagger_0$ corresponds to the IV $pn$-pair creation operator.
$\hat{P}^\dagger_\mu$ creates IS $pn$  
pair where two nucleons couple to $(S=1,T=0)$,
and $\mu$ is the label of a spin component $S_z$.
$\hat{F}^{\mu}_\nu$ are the GT operators.
In addition to these one-body operators, the spin operators
$\hat{S}_\mu$,
the isospin operators
$\hat{T}_\nu$,
and the number operator $\hat{N}$ constitute so(8) algebra. 
When the single-particle-energy term $\hat{h}_{\rm spe}$ is omitted,
all the matrix elements of the Hamiltonian are analytically evaluated
with the basis belonging to the irreducible representation of SO(8) group.
This Hamiltonian is called the SO(8) Hamiltonian \cite{Engel1997,Pang1969,Evans1981}.

\subsection{Proton--neutron-mixing HFB + GCM}
To incorporate the $pn$-pair condensation,
we solve the generalized Hartree--Fock--Bogoliubov (HFB) equation~\cite{Goodman1979},
\begin{equation}
 \left(\begin{matrix}
     h-\lambda & \Delta \cr
     -\Delta^\ast & -h^\ast+\lambda
           \end{matrix}\right)
           \left(\begin{matrix}
     U  \cr
     V
           \end{matrix}\right)
           =
     E
           \left(\begin{matrix}
     U  \cr
     V
           \end{matrix}\right).
\end{equation}
In the generalized HFB equation, the quasiparticle operators are
mixtures of proton and neutron operators, that is, 
\begin{equation}
\alpha^\dagger_i=(U_pc^\dagger_p)_i+(V_pc_p)_i+(U_nc^\dagger_n)_i+(V_nc_n)_i.
\end{equation}
$h$ and $\Delta$ become $2\times2$ matrices in the isospin space.
We solve the generalized HFB equation neglecting Fock terms. 
The solutions of the generalized Hartree--Bogoliubov (HB) equation are used as the basis state of the GCM, and 
the Fock terms are re-introduced when computing the Hamiltonian kernel in the GCM calculation.
Adding the linear constraint terms with Lagrange multipliers,
we obtain the constrained HB states $\ket{P_0,R_0}$ with a positive parity
having the expectation values
$\langle \hat{P}_{0}\rangle=P_0$ and 
$\langle \hat{R}_{0}\rangle/i=R_0$.
\footnote{
Note that in our choice of the phase 
$\langle \hat{P}_0\rangle$ is real and
$\langle \hat{R}_0\rangle$ is pure imaginary.
}
In general,
the generalized HB solutions break the time-reversal and the axial symmetries.
In the following calculation, we impose the axial symmetry by the conditions
$\langle \hat{P}_{\pm1}\rangle =0$.
Here, $\langle \hat{R}_{\pm1}\rangle$ are variationally determined under the proton and neutron number conditions.
Even if the minimum-energy HB solution has $\langle \hat{P}_0\rangle=0$ and $\langle \hat{R}_0\rangle=0$,
the $pn$-pair correlation may play an important role in the beyond-mean-field level,
especially when the potential energy surface (PES) is relatively flat
in the plane of $(P_0,R_0)$.
In order to take into account the beyond-mean-field correlations,
we use a trial wave function in the GCM ansatz \cite{ring} with
the weight functions $f_J(P_0,R_0)$,
\begin{equation}
|\Psi_{JM}\rangle=\int^{R_{\rm max}}_0 dR_0 \int^{P_{\rm max}}_0 dP_0 f_J(P_0,R_0)P^NP^ZP^J_{M0}|P_0,R_0\rangle.
\end{equation}
$P^N$, $P^Z$, and $P^J_{MK}$ are neutron-number, proton-number, and angular-momentum projection operators, respectively.
The values $P_{\rm max}$ and $R_{\rm max}/i$ are the maximum values in the adopted model space.
To the best of our knowledge,
for describing odd--odd systems,
no attempts have been made before
to perform the two-dimensional GCM calculation with both $P_0$ and $R_0$ as
generator coordinates.
It should be noted that odd-particle-number states are constructed as
the one-quasiparticle-excited states in the HFB theory
with the conventional like-particle pairing only.
In contrast, the present $pn$-mixed states contain components with both even and
odd numbers of particles.
We are able to extract the odd-particle-number states
by performing the number projections.

The variation of the energy $\bra{\Psi_{JM}}H\ket{\Psi_{JM}}$
with respect to the weight functions $f_J$ leads to
the Hill--Wheeler equation, 
\begin{equation}
	\int dR'_0dP'_0 \left[\langle P_0,R_0|HP^NP^ZP^J_{M0}|P'_0,R'_0\rangle-E_k\langle P_0,R_0|P^NP^ZP^J_{M0}|P'_0,R'_0\rangle \right]f_{J}^{(k)}(P'_0,R'_0)=0.
\end{equation}
Solving the generalized eigenvalue equation,
we obtain the eigenenergy and the weight function $f_J^{(k)}$ corresponding to
the $k$-th eigenstate with the good quantum numbers $(N,Z)$ and $(J,M)$.
In the numerical calculation,
the integration is discretized with
the mesh sizes of $\Delta P_0=\Delta R_0=0.5$.
For calculations of the Hamiltonian and the norm kernels,
we use 11-points and 16-points Gauss–Legendre quadrature
for the particle-number
and the angular-momentum projections, respectively.
In order to avoid the numerical instability associated with the overcompleteness
of the GCM basis, we exclude the components corresponding to the small eigenvalues
of the norm kernel \cite{ring}.

\section{Magnetic moment and \texorpdfstring{$M1$}{M1} transition \label{sec:M1}}
The magnetic dipole operator $\hat{\bm{\mu}}$ is defined as 
\begin{equation}
\hat{\bm{\mu}} = \hat{\bm{\mu}}_L + \hat{\bm{\mu}}_S
=\mu_N \hat{\bm L}_p
	+\mu_N \left(g_p\hat{\bm S}_p +g_n\hat{\bm S}_n \right),
\end{equation}
where $\mu_N$ is the nuclear magneton,
$\hat{\bm L}_q$ and $\hat{\bm S}_q$ ($q=n,p$) are
the orbital angular momentum and the spin operators, respectively.
$g_q$ ($q=n,p$) are the spin $g$-factor
with $g_p=5.586$ and $g_n=-3.826$.
The nuclear magnetic moment is given by
\begin{equation}
\mu=\langle J,M=J|\hat{\mu}_z|J,M=J\rangle. 
\end{equation}
The $M1$ reduced transition rate between 
the initial $|J_i\rangle$ and the final state $|J_f\rangle$
is 
\begin{equation}
B(M1;J_i\rightarrow J_f)=\frac{1}{2J_i+1}\frac{3}{4\pi}|\langle J_f\|\hat{\mu}\|J_i\rangle|^2 .
\end{equation}
 
For convenience, we decompose $\hat{\bm \mu}$ into the IS and the IV parts,

\begin{equation}
\hat{\bm \mu}=\mu_N \left\{\frac{1}{2}\left[\hat{\bm L}+(g_p+g_n)
\hat{\bm S}\right]
-\frac{1}{2}\left[\check{\bm L}+(g_p-g_n)
\check{\bm S} \right]\right\},
\end{equation}
with $\hat{\bm L}\equiv \hat{\bm L}_n+\hat{\bm L}_p$,
$\hat{\bm S}\equiv \hat{\bm S}_n+\hat{\bm S}_p$,
$\check{\bm L}\equiv \hat{\bm L}_n-\hat{\bm L}_p$,
and $\check{\bm S} = \hat{\bm S}_n-\hat{\bm S}_p$.
The values $g_p - g_n = 9.412$ and $g_p+g_n = 1.76$ indicate
that the IV spin part is dominant.
It corresponds to the $T_z=0$ component of the GT 
operator ($\check{S}_\mu=\hat{F}^\mu_0$),
and the properties of the IV spin $M1$ transition
may be qualitatively applicable to the GT
transition in the charge-exchange reactions and $\beta$ decay.

Utilizing the GCM wave functions based on the axially symmetric basis states,
the reduced $M1$ transition probability is evaluated as \cite{Guzman2002}
\begin{align}
|\langle J_f\|\hat{\mu}\|J_i\rangle|^2=\Bigg|\int dR_0dR'_0dP_0dP'_0 f^*_{J_f}(P'_0,R'_0)f_{J_i}(P_0,R_0)
\langle N,Z,J_f;P'_0,R'_0\|\hat{\mu}\|N,Z,J_i;P_0,R_0\rangle\Bigg|^2,
\end{align}
where
\begin{equation}
\begin{split}
\langle N,&Z,J_f;P'_0,R'_0\|\hat{\mu}\|N,Z,J_i;P_0,R_0\rangle\\
&=\frac{(2J_i+1)(2J_f+1)}{8\pi^2}(-1)^{J_i-1}
 \times\sum_{\nu'}\left(\begin{matrix}
  J_i & 1 &J_f \cr
  -\nu' & \nu'&0
   \end{matrix}\right) \\
 &\quad\times \int_0^{2\pi} d\phi_n\int_0^{2\pi} d\phi_p   \int_{0}^{\pi} d\beta\sin\beta   d^{J_i*}_{-\nu'0}(\beta)\langle P_0', R_0'|\hat{\mu}_{\nu'}e^{-\phi_n(\hat{N}-N)}e^{-\phi_p(\hat{Z}-Z)}e^{-i\beta \hat{J}_y}|P_0, R_0\rangle .
\end{split}   
\end{equation}
Here, $\hat{N}$ and $\hat{Z}$ are the neutron and the proton number operators, respectively, and $\hat{J}_y=\hat{L}_y+\hat{S}_y$.

\section{Results \label{sec:results}}
\subsection{Validity of the \texorpdfstring{$pn$}{pn}-pair GCM for \texorpdfstring{$N=Z$}{N=Z} odd--odd systems \label{sec:SO8}
} 

We first show the validity and usefulness of the $pn$-pair GCM
for a description of odd--odd nuclei,
by comparing the results with those of the exact solution of 
the SO(8) Hamiltonian of the degenerate $l$ shells without $\hat{h}_{\rm spe}$.
In this subsection, we treat $N=Z=5$ nuclei 
with the degeneracy of the two $l$ shells, $\Omega=(2l_1+1)+(2l_2+1)=12$
($p$ and $g$ shells).
The eigenstates of the SO(8) Hamiltonian have the following quantum numbers;
the mass number $A=N+Z$, the total spin $(S,S_z)$, and the isospin $(T,T_z)$.
$A$ and $T_z=(N-Z)/2$ can be replaced by $N$ and $Z$. 
The isospin symmetry is exact but
is broken in the basis wave function of the $pn$-pair GCM.
We do not perform the isospin projection to restore it.
The IS and IV pairing strengths in Eq.~(\ref{Hamiltonian})
are parameterized as
$G_{\rm IS}=g(1-x)/2$ and
$G_{\rm IV}=g(1+x)/2$.
In the SO(8) Hamiltonian, $g$ is the only parameter with the dimension of energy
and determines the energy scale.
The parameter $x$ is varied and the GT interaction $G_{\rm ph}$ is set to zero for simplicity.

\begin{figure}
\centering
\includegraphics[width=8.6cm]{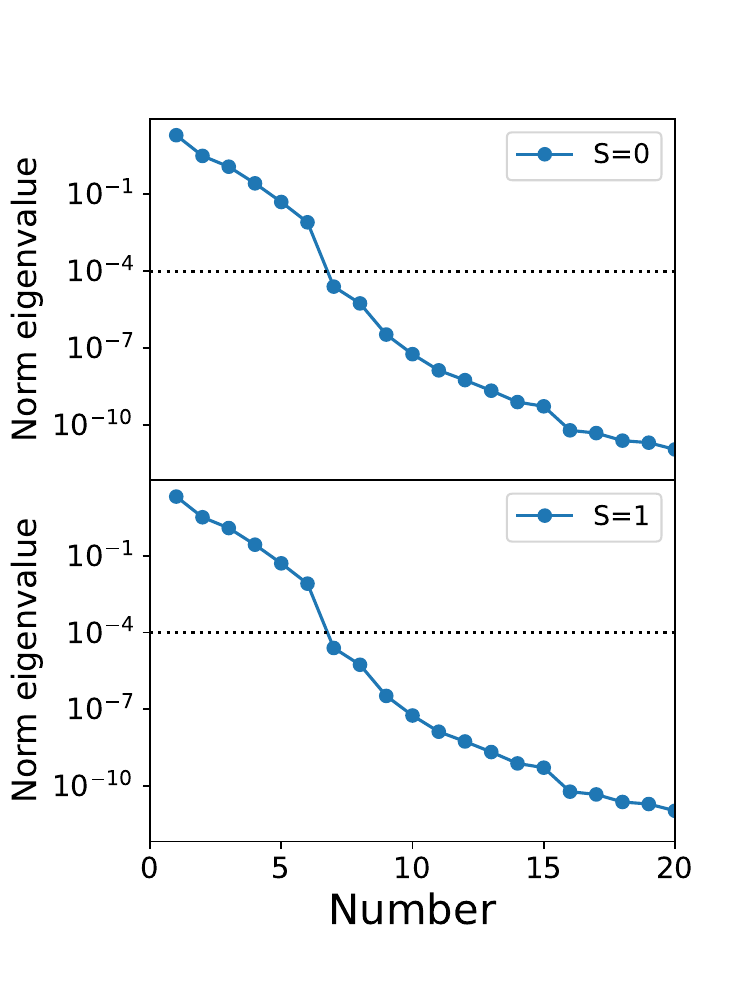}
\caption{Eigenvalues of the norm kernel for the $S=0$ state (the upper panel) and the $S=1$ state (the lower panel).
The horizontal lines indicate the value of the norm cutoff, $10^{-4}$.
}
\label{norm}
\end{figure}

\begin{figure}
\centering
\includegraphics[width=8.6cm]{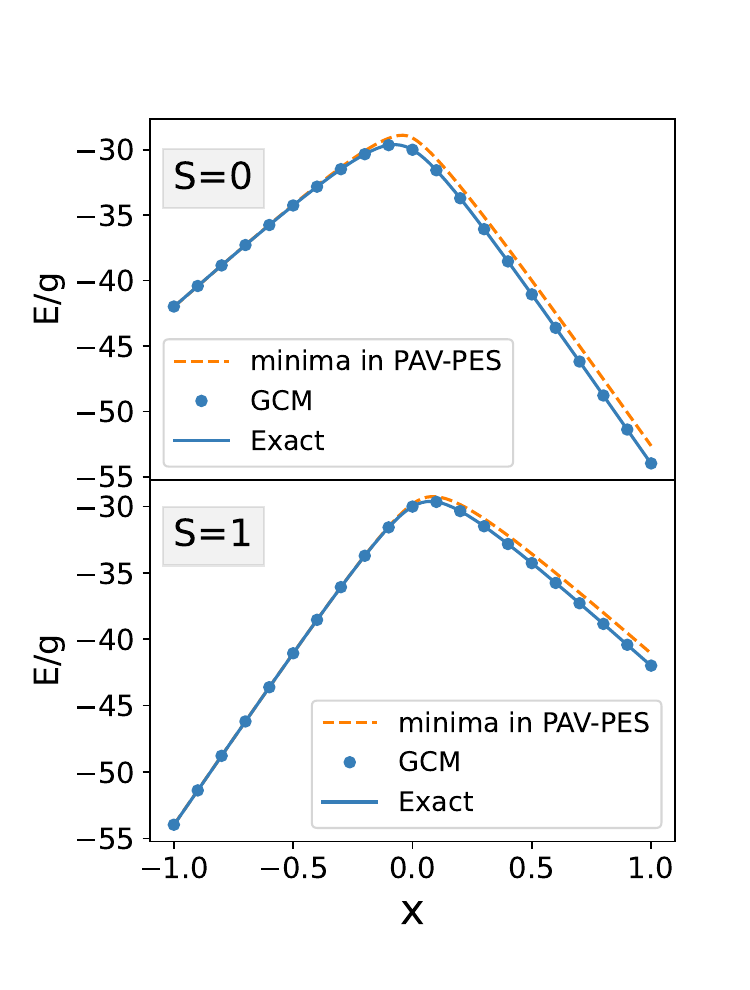}
\caption{Upper panel: Energy of the lowest $S=0$ state for the SO(8) Hamiltonian with $N=Z=5$ and $\Omega=12$ as a function of the parameter $x$.
The dashed line, denoted as ``minima in PAV-PES,''
corresponds to that of the HB solution located at the minima
of the PES in the $(P_0,R_0)$ plane after the particle-number and angular-momentum projections.
The blue circles and solid lines show those of the $pn$-pair GCM and the exact solutions, respectively.
Lower panel: Same as the upper panel except for the lowest $S=1$ state. 
}
 \label{SO8Energy}
\end{figure}
\begin{figure}
\centering
\includegraphics[width=8.6cm]{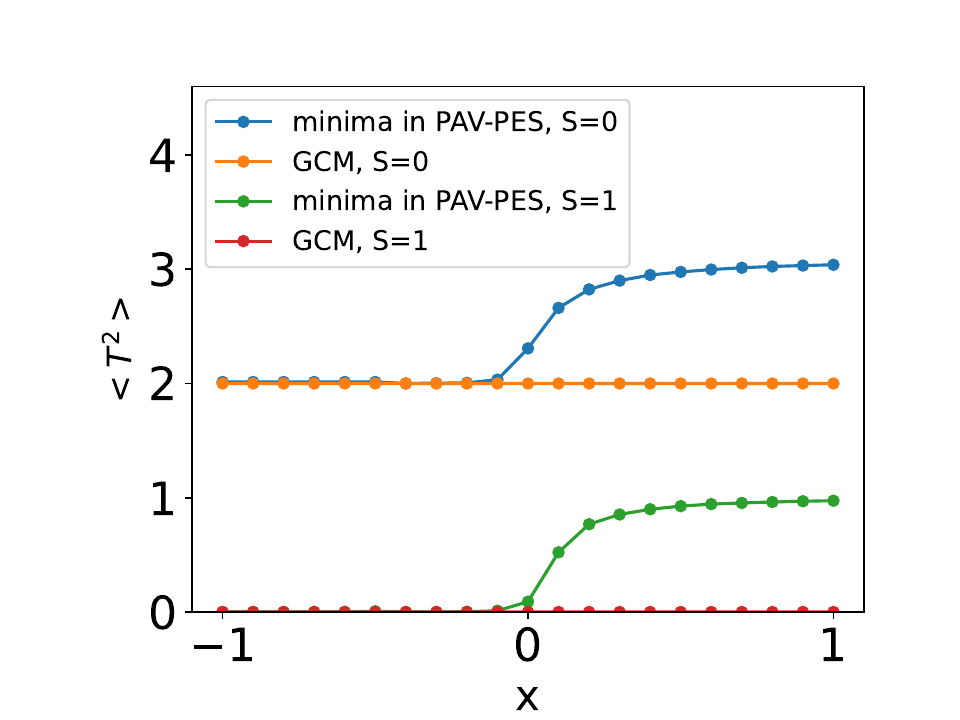}
\caption{
Calculated expectation values of isospin $\hat{T}^2$
for the states obtained with the $pn$-pair GCM and
the minima in PAV-PES states with projection on good $N$, $Z$, and $J=S$.
}
\label{TT}
\end{figure}
\begin{figure}
\centering
\includegraphics[width=8.6cm]{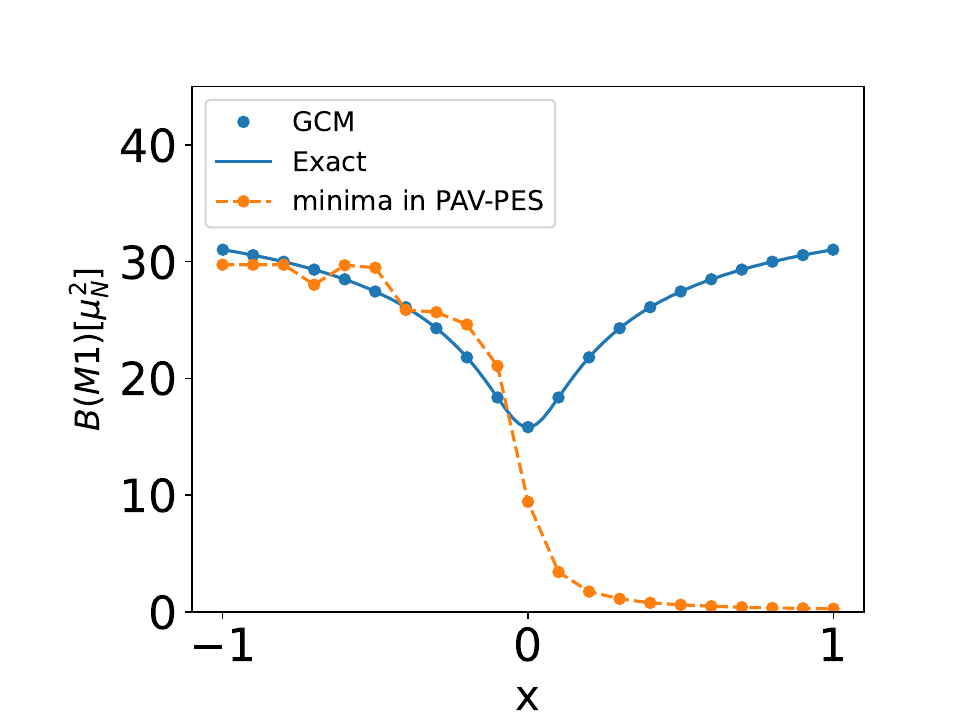}
\caption{$B(M1)$ transition probability from the lowest-energy $S=0$ state to the $S=1$ state obtained as a function of $x$.
The blue circles and the blue line indicate the GCM solution and the exact solutions, respectively.
The orange dashed line shows the result of the PAV-PES states.
}
\label{SO8Transition}
\end{figure}

In Fig. \ref{norm}, the eigenvalues of the norm kernel for the $S=0$ state and the $S=1$ state are plotted in descending order.
The clear gaps are observed around $10^{-4}$; therefore we set the lower limit of the norm eigenvalues to $10^{-4}$.
In the following sections for studies with multi-level models, since we encounter a severe overcompleteness problem, we use a larger value for the norm cutoff.
See section~\ref{sec:18F}.
The energy of the lowest states of $S=0$ and $1$ are shown in Fig.~\ref{SO8Energy}.
We compare the results of the $pn$-pair GCM with the exact ones.
The minimum energy of the projected states $P^NP^ZP^{J=S}_{M0}\ket{P_0,R_0}$ in the $(P_0,R_0)$ plane
is also shown by the dashed line,
which in the followings, we call ``minima in PAV-PES.''
The $pn$-pair GCM almost perfectly reproduces the exact energy,
while there are some deviations in that of the minima in PAV-PES,
especially in $x>0$.
Reference~\cite{Romero2019} shows that the HFB state obtained with
the variation-after-projection (VAP) calculation well reproduces the exact solutions.
Although we perform neither the isospin projection 
nor the VAP for the particle numbers and angular momentum,
the $pn$-pair GCM effectively incorporates full correlations 
for the entire range of the parameter $x$, from $x=-1$ (pure IS pairing)
to $x=1$ (pure IV pairing).
We have confirmed that the isospin symmetry is practically restored
in the $pn$-pair GCM states, e.g., the $S=0$ (1) states have $T=1$ (0)
at high accuracy; see Fig.~\ref{TT}.
The present $pn$-pair GCM has an advantage in its applicability to realistic models.
In addition, the collective wave functions provide an intuitive picture of
the quantum fluctuation in the pairing degrees of freedom.

Next, we analyze the $M1$ transition strength. 
In the SO(8) model, the orbital angular momentum does not contribute to the $M1$ transition and it is sufficient to take the spin part into account only. 
Moreover, 
since the isospin symmetry is exact in the SO(8) Hamiltonian,
the IS transition is prohibited between the $S=0$ ($T=1$) and $S=1$ ($T=0$) states.
Thus, $B(M1)$ is proportional to  $|\langle S=1|\hat{F}^0_0|S=0\rangle|^2$.
Figure~\ref{SO8Transition} shows $B(M1)$ as a function of $x$.
Because SO(8) Hamiltonian has a symmetric form for the exchange of the spin and isospin,
the value of $B(M1)$ is symmetric about $x=0$.
\footnote{
At $x=0$ where the Hamiltonian becomes SU(4) symmetric \cite{Engel1997}, 
$B(M1)$ only between the same SU(4) multiplet 
(Wigner supermultiplet)
becomes finite, for example, between $(S,T)=(0,1)$ and $(1,0)$.

In this case $\langle S=1,T=0|\hat{F}^0_0|S=0,T=1\rangle=1$.
}
The $pn$-pair GCM calculation reproduces the exact $B(M1)$ values very precisely.

The single HB state projected on good $N,Z$, and $J=S$ approximately reproduces
the energy of the lowest states (Fig.~\ref{SO8Energy}).
However, it fails to reproduce the correct $B(M1)$ value as shown
in the orange dashed line in Fig.~\ref{SO8Transition}. 
The deviation is larger in the region of $x>0$ where the IV pairing is dominant.
Together with the observation in Fig.~\ref{TT},
this suggests that the isospin symmetry breaking
may be a main source of the deviation.

The accuracy of the $pn$-pair GCM results suggest that
the GCM with the two-dimensional generator coordinates $(P_0,R_0)$
is suitable for describing the $J=0$ and $J=1$ 
lowest-energy states of the $N=Z$ odd--odd nuclei. 
The $pn$-pair GCM also reproduces properties of low-lying excited states,
which may be useful for the analysis of the collective excitation modes
such as $pn$-pair vibrations. 

\subsection{
Structure and magnetic property in the \texorpdfstring{$N=Z$}{N=Z} odd--odd nuclei \label{sec:18F}
}

\subsubsection{Energy difference between the lowest \texorpdfstring{$J=0$}{J=0} and \texorpdfstring{$J=1$}{J=1} states}
In $N=Z$ nuclei, the protons and neutrons
occupy similar orbits.
Particularly,
the $pn$-pair correlation is expected to play an important role
in odd--odd nuclei where unpaired neutrons and protons exist
in the like-particle pairing.
We apply the $pn$-pair GCM to the $N=Z$ odd--odd nuclei.
For this purpose, the single-particle-energy term,
$\hat{h}_{\rm spe}=\sum_i \varepsilon_i c_i^\dagger c_i$,
that breaks the SO(8) symmetry,
is activated in the Hamiltonian (\ref{Hamiltonian}).
The $jj$-coupling single-particle energies with different values
for protons and neutrons also break the spin and the isospin symmetry.

Among those $N=Z$ nuclei,
$^{18}$F shows a large $M1$ transition strength between the lowest $0^+$ and
the $1^+$ states.
The large $B(M1)$ was reproduced in the three-body model calculation,
assuming an $^{16}$O ${\rm core}+p+n$ structure
with the residual $pn$ interaction
\cite{Tanimura2014}.
The antisymmetrized molecular dynamics (AMD) calculation also
reproduced the $M1$ transition strength, and
the spatial 
behavior of the $pn$ correlation
was analyzed in detail in Ref.~\cite{Kanada2014}.

In the following calculation,
in order to describe the $N=Z$ $sd$-shell nuclei including $^{18}$F,
the $s+p+sd$ shells are used for the single-particle model space.
We use the canonical single-particle energies of $^{16}$O
for the single-particle energies $\varepsilon_i$ of $^{18}$F and
those of $^{24}$Mg for $\varepsilon_i$ of $^{22}$Ne and $^{26}$Al. 
The canonical single-particle energies are obtained
with the Skyrme SkP energy density functional \cite{Dobaczewski1984}.
We solve the HFB equation using the {\sc hfbtho} (v1.66p) code \cite{STOITSOV200545} within the spherical symmetry.
The obtained single-particle energies are summarized in Table~\ref{s.p.e.}.
Although the $d_{3/2}$ orbitals are unbound,
for a proper description of the IS pairing correlations,
it is important to include those resonance states
in the model space as the spin-orbit partner of $d_{5/2}$ \cite{Tanimura2014}.
We fix the IV pairing strength $G_{\rm IV}=1$~MeV 
that is determined to reproduce the empirical pairing gap of $^{26}$Ne.
The GT interaction strength is set to $G_{\rm ph}=20.8/A^{0.7}$~MeV
to reproduce the position of the energy peak of the GT 
giant resonance of $^{48}$Ca, $^{90}$Zr, and $^{208}$Pb
in the $pn$-QRPA calculation \cite{Homma1996}.
We examine the role of the IS pairing,
by changing the IS pairing strength $G_{\rm IS}$.
In this subsection, instead of the $x$ parameter,
we introduce a parameter $g_{\rm pp}$
which is the ratio between the IS pairing and the IV pairing strengths,
$g_{\rm pp}=G_{\rm IS}/G_{\rm IV}$.

In the following calculation, we set the value of the norm cutoff, $0.65$, which results in the collective dimension (the number of the norm eigenvalues larger than the cutoff) equal to 8
for states projected on $J=0$ and 1 in $^{18}$F, $^{22}$Ne, and $^{26}$Al.
The reason for this choice is as follows.
We have found that,
when we increase the collective dimension by reducing the value of the norm cutoff, the calculated collective wavefunction $g^J(P_0, R_0)$ \cite{ring} exhibits a double peak structure even for the ground state.
Since the diagonal elements of the Hamiltonian kernel do not show multiple local minima,
this is most probably due to the overcompleteness of the adopted collective space.

\begin{table}
\caption{Single-particle energies $\varepsilon_i$ for $^{18}$F, $^{22}$Ne, and $^{26}$Al in the unit of MeV.}
    \centering
    \begin{tabular}{lcccccc}
        \hline 
         & $1s_{1/2}$ & $1p_{3/2}$ & $1p_{1/2}$ & $1d_{5/2}$ & $2s_{1/2}$ &$1d_{3/2}$\\
        \hline 
        proton ($^{18}$F)     & $-25.55$        & $-14.91$      & $-10.35$   & $-3.63$     &  $-0.65$   & 4.99   \\
        neutron ($^{18}$F)     & $-29.02$  & $-18.27$ & $-13.61$ & $-6.84$    & $-3.83$   &1.98   \\
        \hline 
        proton ($^{22}$Ne and $^{26}$Al)     & $-29.79$        & $-18.90$      & $-14.52$   & $-8.38$     &  $-6.22$   & $-0.65$   \\
        neutron ($^{22}$Ne and $^{26}$Al)     & $-34.74$  & $-23.74$ & $-19.29$ & $-13.05$    & $-10.84$   & $-5.13$   \\
        \hline 
    \end{tabular}
    \label{s.p.e.}
\end{table}

\begin{figure}
\centering
\includegraphics[width=15.05cm]{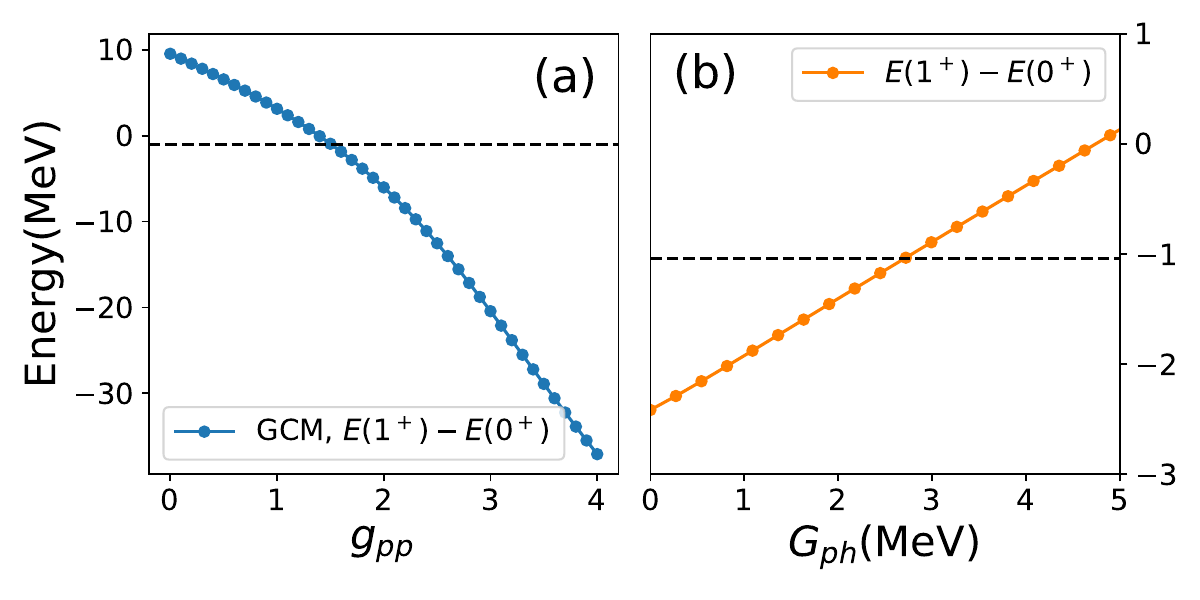}
\caption{
Energy difference between the $1_1^+$ and the $0_1^+$ states in $^{18}$F obtained with the $pn$-pair GCM as a function of $g_{\rm pp}$ (the left panel) or $G_{\rm ph}$ (the right panel).
$G_{\rm ph}$ and $g_{\rm pp}$ are fixed to be $2.72$~MeV and $1.51$ in the left and right panels, respectively.
The dashed line shows the experimental value.}
\label{18FEnergy}
\end{figure}

The calculated energy difference between $1_1^+$ and $0_1^+$ states is
plotted in Fig.~\ref{18FEnergy}.
The energy difference $E(1_1^+)-E(0_1^+)$ decreases as a function of $g_{\rm pp}$.
This is because, even though the spin symmetry is partially broken,
the IS $(S=1, T=0)$ pair correlation is stronger in the $J^\pi=1^+$
state than in the $0^+$ state. 
The clear correlation in Fig.~\ref{18FEnergy}
suggests that the energy difference $E(1_1^+)-E(0_1^+)$ may serve
as a possible observable to 
determine the values of $g_{\rm pp}$ in the Hamiltonian.
The experimental value in $^{18}$F,
$E(1^+_1)-E(0^+_1)=-1.04$~MeV,
leads to $g_{\rm pp}=1.51$.
This ratio is consistent with the estimated value in the previous
studies~\cite{Cohen1965}.
On the other hand, the energy difference is less sensitive to the GT interaction strength $G_{\rm ph}$.
The same behaviors of the $E(1_1^+)-E(0_1^+)$ are obtained
for the $^{22}$Na and $^{26}$Al.
The calculated energy differences are given in Table~\ref{tableI}.
Although the agreement with experiments is not quantitatively perfect,
they are qualitatively well reproduced
with the common interaction parameters: $G_{\rm IV}=1$~MeV, $g_{\rm pp}=1.51$, and $G_{\rm ph}=20.8/A^{0.7}$~MeV.

\begin{table}[hbtp]
 \caption{The energy differences between the $0_1^+$ and $1_1^+$ states, $E(1_1^+)-E(0_1^+)$ in $^{18}$F, $^{22}$Na, and $^{26}$Al. 
 The interaction strengths are $G_{\rm IV}=1$~MeV, $g_{\rm pp}=1.51$, and $G_{\rm ph}=20.8/A^{0.7}$~MeV.
\label{tableI}
}
\centering
  \begin{tabular}{ld{5.5}d{2.8}}
    \hline
& \multicolumn{1}{c}{Exp. (MeV)}
& \multicolumn{1}{c}{$pn$-pair GCM (MeV)}\\
  \hline
  $^{18}$F & -1.04 & -1.04 \\  
  $^{22}$Na& -0.07 & -0.07 \\
  $^{26}$Al& 0.83 &  1.97 \\ 
    \hline
  \end{tabular}
\end{table}
\begin{figure}
\centering
\includegraphics[width=8.6cm]{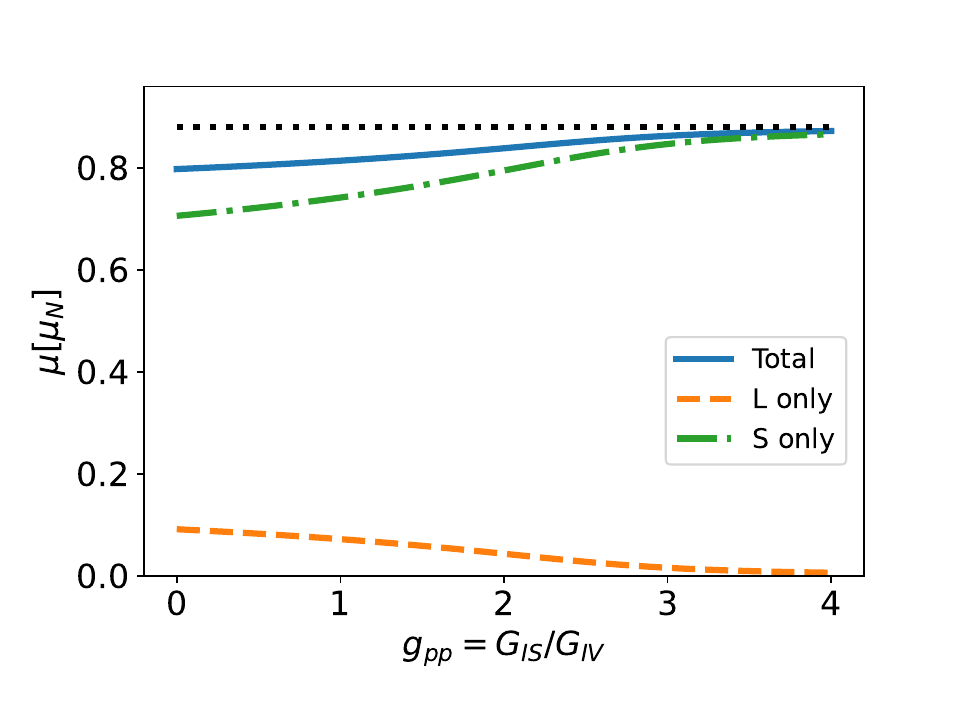}
\caption{Magnetic moment $\mu$ of $1^+_1$ states in $^{18}$F as a function of $g_{\rm pp}$. 
The blue solid line shows the total magnetic moment. The orange dashed and the green dash-dotted lines show the orbital and the spin parts, respectively.
The black dotted line indicates $\mu=(g_p+g_n)/2=0.88 \mu_N$.
}
\label{MM18F}
\end{figure}

\subsubsection{Magnetic moment and \texorpdfstring{$M1$}{M1} transition strengths}

In Fig.~\ref{MM18F}, we plot the magnetic moment of the $1_1^+$ state
in $^{18}$F as a function of $g_{\rm pp}$.
With a larger value of $g_{\rm pp}$, the IS pair configuration becomes dominant. 
The magnetic moment reaches the IS spin $g$-factor value $(g_p+g_n)\mu_N/2=0.88\mu_N$. 
An IS pair of the valence proton and neutron $(S=1,T=0)$
dominantly contributes to the magnetic moment.
A partial restoration of the SO(5) symmetry,
which becomes exact at the pure IV-pair model ($g_{\rm pp}=\infty$),
takes place by the strong IS pairing strength.

\begin{table}[hbtp]
 \caption{Magnetic moment $\mu$,
its orbital part $\mu_L$, and the spin part $\mu_S$ in units of $\mu_N$ in $1^+_1$ states.
The interaction strengths are $G_{\rm IV}=1$~MeV, $g_{\rm pp}=1.51$, and $G_{\rm ph}=20.8/A^{0.7}$~MeV.
\label{tableII}
}
\centering
  \begin{tabular}{ccccc}
    \hline
  &Exp. & $\mu$ & $\mu_L$ & $\mu_S$ \\
  \hline
  $^{18}$F&&0.825&0.059&0.767\\  
  $^{22}$Na&$0.535\pm0.10$&0.832&0.056&0.776\\
  $^{26}$Al&&0.834 &0.058 &0.777 \\ 
    \hline
  \end{tabular}
\end{table}
The obtained values of the magnetic moment
in $^{18}$F, $^{22}$Na, and $^{26}$Al
are summarized in Table~\ref{tableII}.  
Although 
the magnetic moment of $1^+_1$ state in $^{18}$F has not been measured,
the present value well agrees with the value $0.834\mu_N$ obtained
with the three-body model \cite{Tanimura2014}.
The $pn$-pair GCM overestimates the magnetic moment for $^{22}$Na,
$\mu_{\rm exp}=0.535 \mu_N$.
The correlations missing in the present $pn$-pair GCM calculation,
such as the quadrupole deformation and the $pn$ pair that couples to $(L,S,T)=(2,1,0)$,
may be responsible for the discrepancy.

\begin{figure}
\centering
\includegraphics[width=15.05cm]{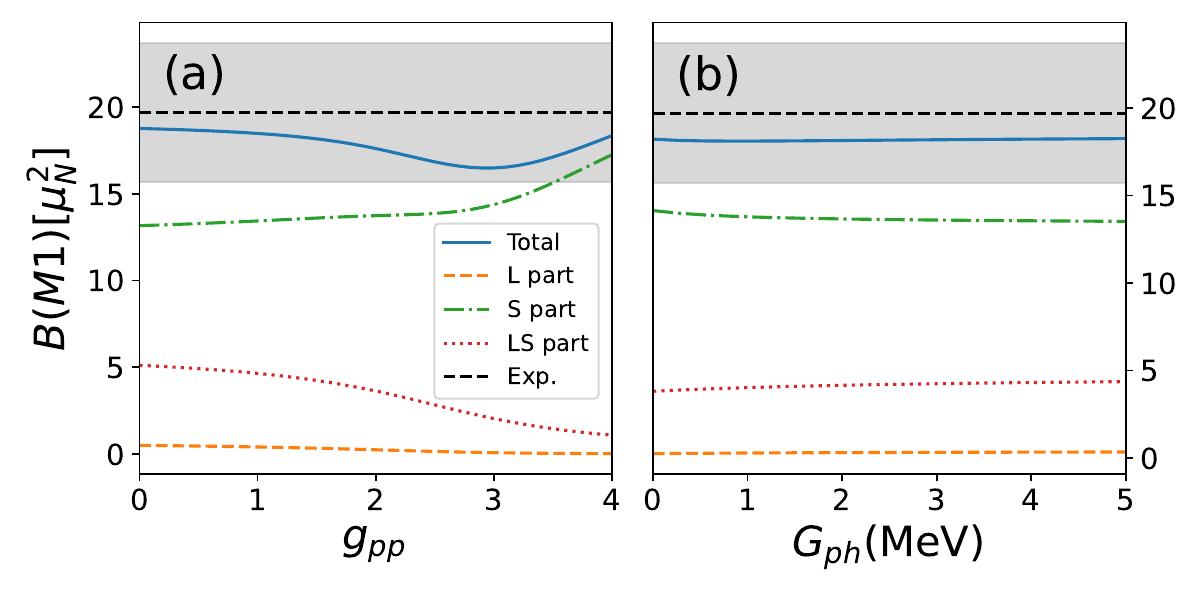}
\caption{The reduced $M1$ transition probability, $B(M1;0_1^+\rightarrow 1_1^+)$ in $^{18}$F as a function of $g_{\rm pp}$ (the left panel) or $G_{\rm ph}$ (the right panel). 
The blue solid line shows the total $M1$ transition probability.
The orange dashed line and the green dash-dotted line show the contribution from the orbital part $B(M1)_L$ or spin part $B(M1)_S$, respectively. 
The interference term $B(M1)_{LS}$ is shown in the red dotted line.
The black dashed line with the shaded region shows the experimental data and its error.
\label{BM1F}
}
\end{figure}

Next, we present the $M1$ transition strengths
between the $1_1^+$ and $0_1^+$ states.
We plot the reduced $M1$ transition probability $B(M1)$ as a function of $g_{\rm pp}$ and $G_{\rm ph}$ in Fig.~\ref{BM1F}(a) and (b), respectively.
In addition to the total transition probability, we decompose the contributions from each term, the spin $M1$ strength $B(M1)_S=\frac{3}{4\pi}|\langle 1_1^+||\hat{\mu}_S||0_1^+\rangle|^2$, the orbital $M1$ strength $B(M1)_L=\frac{3}{4\pi}|\langle 1_1^+||\hat{\mu}_L||0_1^+\rangle|^2$ and the interference term $B(M1)_{LS}=\frac{6}{4\pi}{\rm Re}(\langle 1_1^+||\hat{\mu}_L||0_1^+\rangle\langle 0_1^+||\hat{\mu}_S||1_1^+\rangle)$.
We find that the $B(M1)$ value gradually decreases as a function of
$g_{\rm pp}$ up to $g_{\rm pp}\lesssim 3$, then, increases beyond that.
In the range of realistic values of $g_{\rm pp}<2$,
$B(M1)$ may not be a good probe of the IS pairing strength.
On the other hand, $B(M1)$ is less sensitive to the $G_{\rm ph}$, similarly to the energy difference in Fig. \ref{18FEnergy}.

\begin{figure}
\centering
\includegraphics[width=12.9cm]{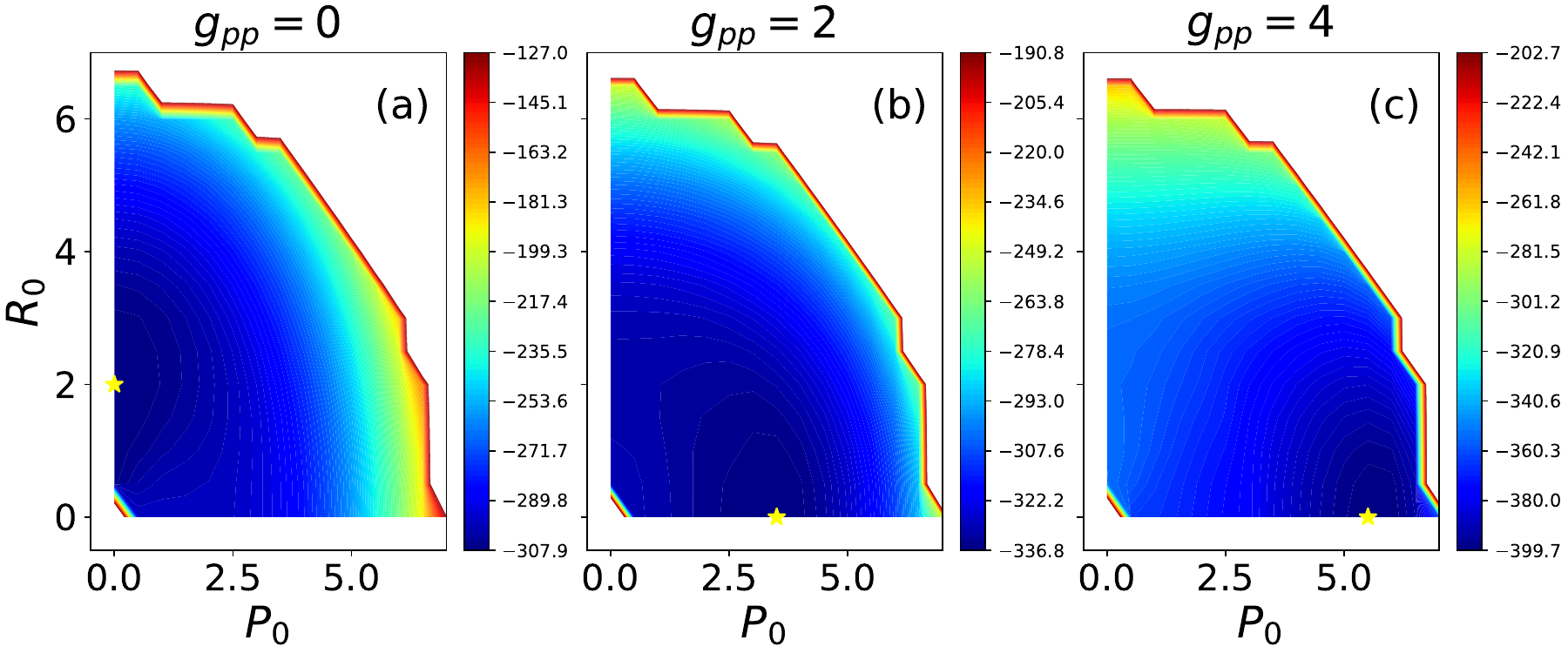}
\caption{
PES without the angular momentum projection in the unit of MeV,
in the two-dimensional $(P_0,R_0)$ plane for $^{18}$F.
The number projection is performed.
The $g_{\rm pp}$ values are 0, 2, and 4 for the left, the center
and the right panels, respectively.
The star represents the energy minima in each graph.
$P_0=R_0=0$ solution does not have $N=Z=9$ component; therefore the origins are excluded in these plots.
}
\label{PESnoJ}
\end{figure}
\begin{figure}
\centering
\includegraphics[width=12.9cm]{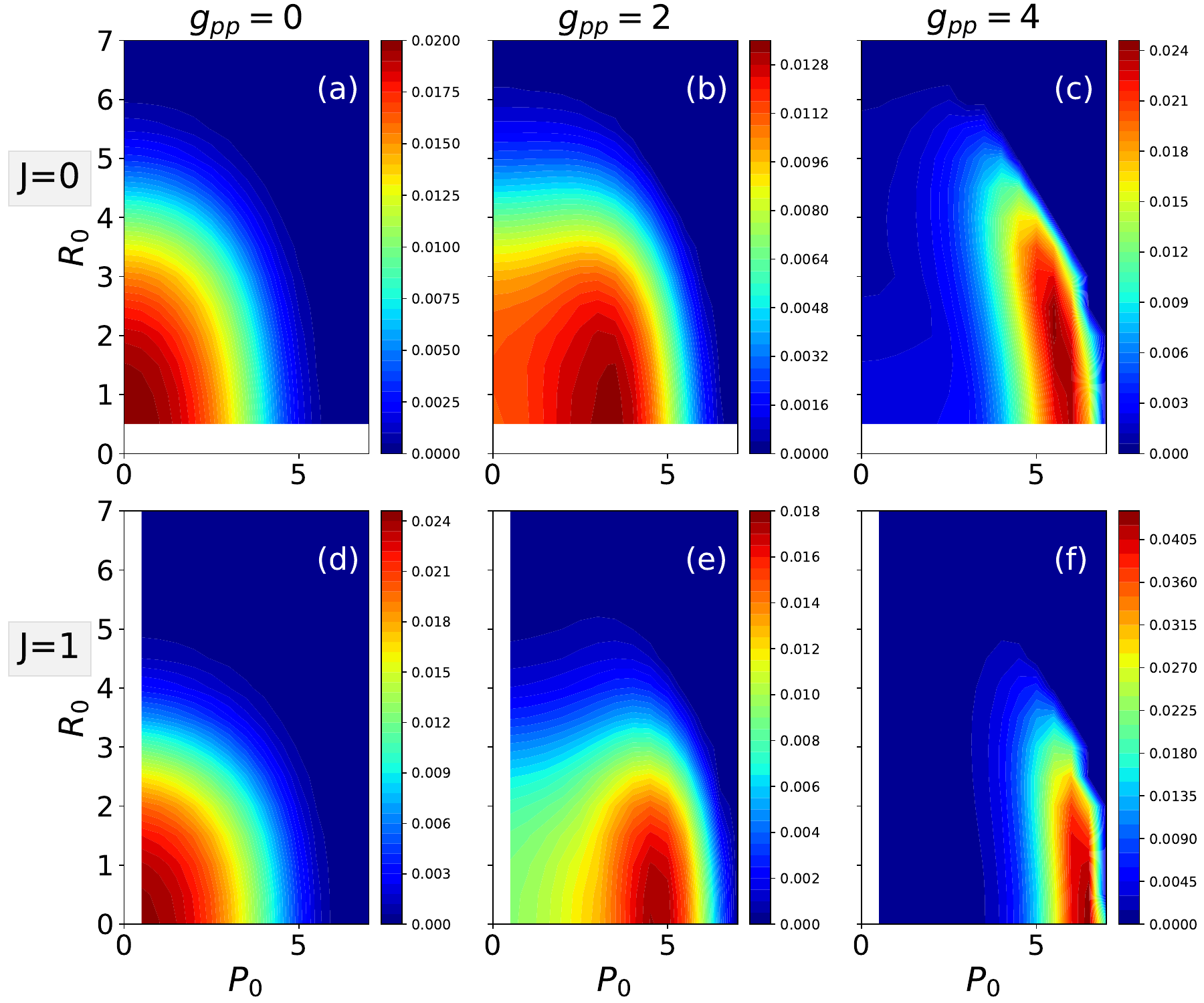}
\caption{
Square of the collective wave functions for $^{18}$F.
Upper panels: Those of the $0^+$ states, $|g^{J=0}(P_0,R_0)|^2$.
Lower panels: Those of the $1^+$ states, $|g^{J=1}(P_0,R_0)|^2$.
The $g_{\rm pp}$ values are 0, 2, and 4 for
the left, the center, and the right panels, respectively.
Note that the projection on $J=0$ (1) produces no state at $R_0=0$ ($P_0=0$).
}
\label{gwave}
\end{figure}

In the GCM framework, in addition to the experimental observables such as the excitation energies and the transitions, one can visualize the collective wave function
in the collective space,
the $(P_0,R_0)$ space in the present case.
We analyze the interplay between the IS and IV $pn$-pairing correlations
on the magnetic property in these odd--odd nuclei.
The PESs defined as $\langle P_0,R_0|HP^NP^Z|P_0,R_0\rangle/\langle P_0,R_0|P^NP^Z|P_0,R_0\rangle$ are shown in Fig.~\ref{PESnoJ}, and the collective wave functions $g^{J=0}(P_0,R_0)$ and $g^{J=1}(P_0,R_0)$ 
are shown in Fig.~\ref{gwave}.

In the case that the IS pairing is switched off
[Fig.~\ref{PESnoJ}(a) and Fig.~\ref{gwave}(a) or (d)], 
the energy minimum is located at $P_0=0$, and the collective wave functions also localize around $P_0=0$.
In Fig.~\ref{PESnoJ}(b) for the case with $g_{\rm pp}=2$, the PES has the shallow minimum around $P_0 =5$.
Corresponding to it, the collective wave functions have the peak around the minimum, though it is widely spread in the $P_0$ direction.
In Fig.~\ref{PESnoJ}(c), the PES has a well-developed energy minimum around $P_0 = 6$ 
indicating a complete transition to the IS-pair condensation phase.
In Fig.~\ref{gwave}(c) or (f), the collective wave functions are also localized in the same region.
The prominent increase of $P_0\approx 6$ configurations
in the $g_{\rm pp}>3$ is reflected to the increases of $B(M1)_{S}$ in Fig.~\ref{BM1F}.
In the present model, $g_{\rm pp}=1.51$ is the value that reproduces the energy
difference between $0_1^+$ and $1_1^+$ states in $^{18}$F.
$g_{\rm pp}=4$ is too strong to explain the experimental data.
However, we should note that there is a possibility that
the realistic $N=Z$ nuclei are located close to the critical point
of the IS-pair condensation \cite{Yoshida2014},
which may be probed by the low-energy peak in the GT transition strength
in $^{18}$O($^{3}$He, $t$)$^{18}$F \cite{Fujita2019}.

\begin{figure}
\centering
\includegraphics[width=8.6cm]{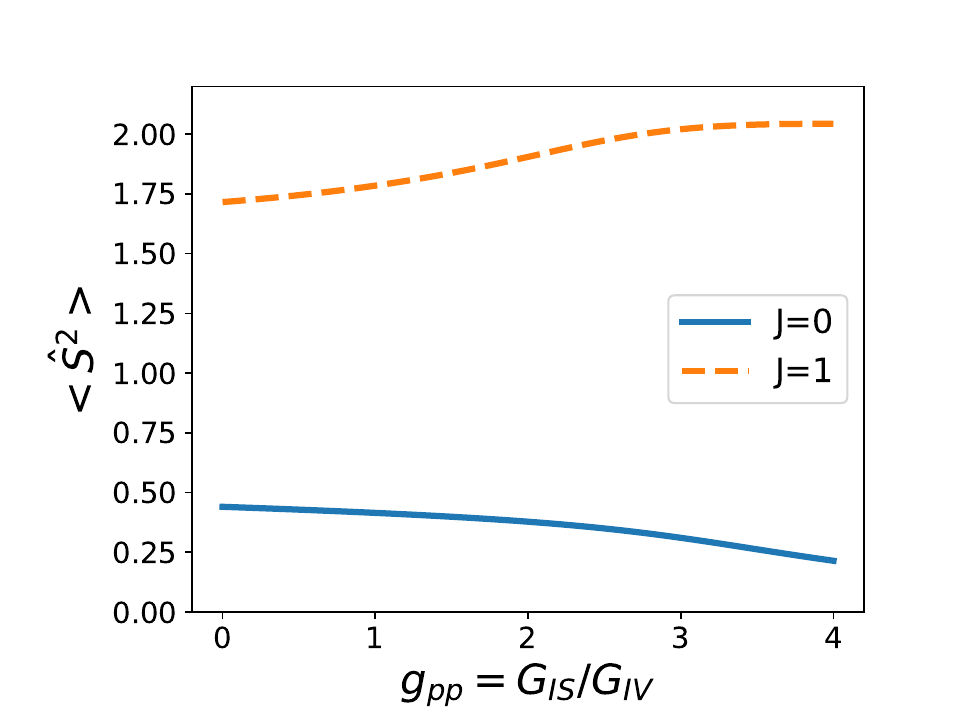}
\caption{Calculated $\langle \hat{S}^2 \rangle $ as a function of $g_{\rm pp}$.
The blue solid and the orange dashed lines correspond to the $J=0$ and $J=1$ states, respectively. 
}

\label{SSFF18F}
\end{figure}

The concave behavior of $B(M1)$ as a function of $g_{\rm pp}$ in Fig.~\ref{BM1F}
is due to the decrease of the interference term $B(M1)_{LS}$,
which is associated with a monotonic reduction of $B(M1)_L$,
from $B(M1)_L=0.42 \mu_N^2$ at $g_{\rm pp}=0$ to $0.01 \mu_N^2$
at $g_{\rm pp}=4$.
At large values of $g_{\rm pp}$, 
since the strong IS $S=1$ pairing configuration becomes dominant, 
we expect that the angular momentum of the $J=1$ state is mainly
carried by the spin.
This explains the reduction of $B(M1)_L$.
The spin dominance is
confirmed by $\langle \hat{S}^2 \rangle $, shown in Fig.~\ref{SSFF18F}.
Increasing $g_{\rm pp}$, the expectation value
$\langle \hat{S}^2 \rangle $ becomes closer to the eigenvalues,
$S(S+1)=0$ and $2$ for $J=0$ and $1$ states, respectively.
The single-particle energies with the spin-orbit splitting
explicitly break the spin symmetry, however,
the spin-SO(5) symmetry is approximately realized at large $g_{\rm pp}$.

\begin{table}[hbtp]
 \caption{
\label{tableIII}
The reduced $M1$ transition probability $B(M1;0^+_1\rightarrow 1^+_1)$
for the $N=Z$ odd--odd nuclei in units of $\mu_N^2$. The experimental values are taken from Refs.~\cite{Lisetskiy1999,TILLEY19951}, among which those of $^{22}$Na and $^{26}$Al are summed values over several transitions.
The calculated total $B(M1)$ is decomposed into the orbital ($B(M1)_L$),
the spin ($B(M1)_S$), and the interference ($B(M1)_{LS}$) contributions.
The interaction strengths are $G_{\rm IV}=1$~MeV,
$g_{\rm pp}=1.51$, and $G_{\rm ph}=20.8/A^{0.7}$~MeV.
 }
  \centering
  \begin{tabular}{cccccc}
    \hline
	  &Exp.& $B(M1)$ & $B(M1)_L$ & $B(M1)_S$ & $B(M1)_{LS}$\\
  \hline
  $^{18}$F&$19.71\pm4.0$&18.17&0.32&13.61& 4.23\\  
  $^{22}$Na&$>12.5$&16.01&0.13&13.24&2.63\\
  $^{26}$Al&$9.4\pm3.0$&14.45 &0.02 &13.26&1.16\\
    \hline
  \end{tabular}
\end{table}

\begin{figure}
\centering
\includegraphics[width=8.6cm]{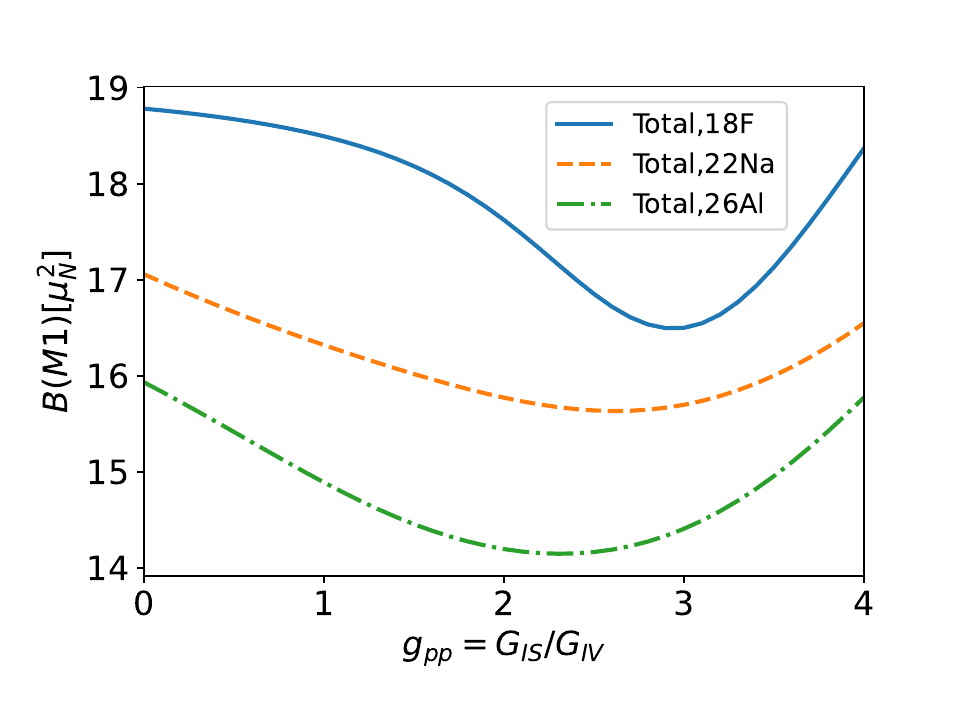}
\caption{
The reduced $M1$ transition probability as a function of $g_{\rm pp}$. The blue solid, orange dashed, and green dash-dotted lines correspond to $B(M1)$ in $^{18}$F, $^{22}$Na, and $^{26}$Al, respectively.
The interaction strengths are $G_{\rm IV}=1$~MeV, $g_{\rm pp}=1.51$, and $G_{\rm ph}=20.8/A^{0.7}$~MeV.}
\label{BM1all}
\end{figure}

We compare the $M1$ transition strength in $^{18}$F, $^{22}$Na, and $^{26}$Al.
Their $g_{\rm pp}$ dependence is shown in Fig.~\ref{BM1all}.
All the lines decrease in small values of $g_{\rm pp}$,
then, increase in large $g_{\rm pp}$.
The magnitude of variation is the largest for $^{26}$Al and the smallest
for $^{18}$F, reflecting the number of valence particles.
The variation originates from the decline of $B(M1)_{LS}$
and the increases of $B(M1)_{S}$.
The greater the number of the valence particles is,
the smaller the $g_{\rm pp}$ value at the minimum of $B(M1)$ becomes.
Therefore, their relative values may possibly provide
information about the IS pairing strength.
The values of $B(M1)$ with $g_{\rm pp}=1.51$ are also summarized in Table~\ref{tableIII}.
The $B(M1)$ value in $^{18}$F is consistent with the experimental values within the standard error. 
For $^{22}$Na and $^{26}$Al, 
the decreasing trend of the $B(M1)$ as a function of the mass number
is reproduced.

\section{Summary and Conclusions \label{sec:summary}}

We have developed the $pn$-pair GCM for describing quantum states of odd--odd nuclei based on the IS and IV $pn$-pair amplitudes as the generator coordinates $(P_0,R_0)$,
combined with the particle-number and the angular-momentum projections.
To assess the feasibility of the method,
we have examined the GCM calculations 
for the spin 0 and 1 states of an odd--odd nucleus
using the SO(8) Hamiltonian
in which the exact solutions are available.
The GCM solutions reproduce both the energy and the $M1$ transition
properties over the entire range of the parameters
between the pure IV and the pure IS pairing limits.
On the other hand, the single HB state with the projections fails to
reproduce the $M1$ transitions.

We have calculated $N=Z$ odd--odd nuclei, $^{18}$F, $^{22}$Na, and $^{26}$Al,
introducing the single-particle-energy term into the Hamiltonian,
with six spherical orbits ($s$, $p$, and $sd$ shells) both for protons and neutrons.
A strong correlation between the energy differences $E(1_1^+)-E(0_1^+)$ 
and the IS $pn$ pairing strength is found.
The IS pairing strength is determined by fitting the energy difference in $^{18}$F.
The same Hamiltonian well reproduces the energy differences
in $^{22}$Na and $^{26}$Al systematically.

Increasing the IS pairing strength,
the calculated PES in the $(P_0,R_0)$ plane
becomes flatter and eventually leads to
the IS pair condensation.
In between the IS and the IV limits,
where most probably the real $N=Z$ nuclei are situated,
the $pn$-pair GCM calculation predicts
the coexistence of the different phases due to beyond-mean-field correlations.

The calculated $M1$ transition strength in $^{18}$F is not so sensitive to
the IS pairing strength.
Increasing the number of the valence nucleons in the $sd$ shell,
the sensitivity grows.
However, the calculated $B(M1)$ does not show a monotonic behavior
as a function of the IS pairing.
It shows a decreasing trend in the small IS pairing regime, then
changes into an increasing behavior in the stronger IS pairing regime.
In order to fix the IS pairing strength,
the experimental information on the $B(M1)$ values
for different $N=Z$ odd--odd nuclei is valuable.

In the present work,
we focus our studies on the role of the IS and IV $pn$ pairings
in odd--odd $N=Z$ nuclei.
Performing the GCM calculations with other degrees of freedom,
such as the quadrupole deformation,
would be an important extension
to understand beyond-mean-field dynamics of both $N=Z$ and $N\ne Z$ nuclei.
Some deviations for the $pn$-pair transfer reaction
between experiments and calculations of the particle-particle
random-phase approximation \cite{Yoshida2014} were reported in
Ref.~\cite{Frauendorf2014}.
It may be interesting to apply
the $pn$-pair GCM to the description of nuclei close to the critical point
and to investigate properties of the $pn$-pair transfer.

\section*{Acknowledgments}
This work was supported by the JSPS KAKENHI
Grants No. JP19KK0343, No. JP20K03964, No. JP22H04569, JP23H01167, and No. JP23KJ1212 
and by JST ERATO Grant No. JPMJER2304.
Numerical calculations were performed using Wisteria/BDEC-01 Odyssey
(the University of Tokyo) provided by the
Multidisciplinary Cooperative Research Program in
the Center for Computational Sciences, University of Tsukuba.

\bibliographystyle{ptephy}
\bibliography{m1}

\begin{thebibliography}{10}

\bibitem{superfluidity}
D.~M. Brink and R.~A. Broglia,
\newblock {\em {Nuclear Superfluidity, Pairing in Finite Systems}},
\newblock  (Cambridge University Press, Cambridge, UK, 2005).

\bibitem{Frauendorf2014}
S.~Frauendorf and A.~O. Macchiavelli, Prog. Part. Nucl. Phys., {\bf 78}, 24 --
  90 (2014).

\bibitem{Macchiavelli2000}
A.~O. Macchiavelli, P.~Fallon, R.~M. Clark, M.~Cromaz, M.~A. Deleplanque, R.~M.
  Diamond, G.~J. Lane, I.~Y. Lee, F.~S. Stephens, C.~E. Svensson, K.~Vetter,
  and D.~Ward, Phys. Rev. C, {\bf 61}, 041303(R) (2000).

\bibitem{Yoshida2014}
Kenichi Yoshida, Phys. Rev. C, {\bf 90}, 031303(R) (2014).

\bibitem{Fujita2014}
Y.~Fujita, H.~Fujita, T.~Adachi, C.~L. Bai, A.~Algora, G.~P.~A. Berg, P.~von
  Brentano, G.~Col\`o, M.~Csatl\'os, J.~M. Deaven, E.~Estevez-Aguado,
  C.~Fransen, D.~De~Frenne, K.~Fujita, E.~Ganio\ifmmode~\breve{g}\else
  \u{g}\fi{}lu, C.~J. Guess, J.~Guly\'as, K.~Hatanaka, K.~Hirota, M.~Honma,
  D.~Ishikawa, E.~Jacobs, A.~Krasznahorkay, H.~Matsubara, K.~Matsuyanagi,
  R.~Meharchand, F.~Molina, K.~Muto, K.~Nakanishi, A.~Negret, H.~Okamura, H.~J.
  Ong, T.~Otsuka, N.~Pietralla, G.~Perdikakis, L.~Popescu, B.~Rubio, H.~Sagawa,
  P.~Sarriguren, C.~Scholl, Y.~Shimbara, Y.~Shimizu, G.~Susoy, T.~Suzuki,
  Y.~Tameshige, A.~Tamii, J.~H. Thies, M.~Uchida, T.~Wakasa, M.~Yosoi, R.~G.~T.
  Zegers, K.~O. Zell, and J.~Zenihiro, Phys. Rev. Lett., {\bf 112}, 112502
  (2014).

\bibitem{Fujita2019}
H.~Fujita, Y.~Fujita, Y.~Utsuno, K.~Yoshida, T.~Adachi, A.~Algora,
  M.~Csatl\'os, J.~M. Deaven, E.~Estevez-Aguado, C.~J. Guess, J.~Guly\'as,
  K.~Hatanaka, K.~Hirota, R.~Hutton, D.~Ishikawa, A.~Krasznahorkay,
  H.~Matsubara, F.~Molina, H.~Okamura, H.~J. Ong, G.~Perdikakis, B.~Rubio,
  C.~Scholl, Y.~Shimbara, G.~S\"usoy, T.~Suzuki, A.~Tamii, J.~H. Thies,
  R.~G.~T. Zegers, and J.~Zenihiro, Phys. Rev. C, {\bf 100}, 034618 (2019).

\bibitem{Bai2014}
C.~L. Bai, H.~Sagawa, G.~Col\`o, Y.~Fujita, H.~Q. Zhang, X.~Z. Zhang, and F.~R.
  Xu, Phys. Rev. C, {\bf 90}, 054335 (2014).

\bibitem{Chazono2021}
Yoshiki Chazono, Kenichi Yoshida, Kazuki Yoshida, and Kazuyuki Ogata, Phys.
  Rev. C, {\bf 103}, 024609 (2021).

\bibitem{Tanimura2014}
Y.~Tanimura, H.~Sagawa, and K.~Hagino, Prog. Theor. Exp. Phys., {\bf 2014},
  053D02 (2014).

\bibitem{Sagawa2016}
H.~Sagawa, T.~Suzuki, and M.~Sasano, Phys. Rev. C, {\bf 94}, 041303 (2016).

\bibitem{Yoshida2021}
Kenichi Yoshida, Phys. Rev. C, {\bf 104}, 014309 (2021).

\bibitem{Jokiniemi2023}
Lotta Jokiniemi and Javier Men\'endez, Phys. Rev. C, {\bf 107}, 044316 (2023).

\bibitem{ring}
P.~Ring and P.~Schuck,
\newblock {\em The Nuclear Many-Body Problem},
\newblock  (Springer-Verlag, Berlin, 2000).

\bibitem{Engel1997}
J.~Engel, S.~Pittel, M.~Stoitsov, P.~Vogel, and J.~Dukelsky, Phys. Rev. C, {\bf
  55} (1997).

\bibitem{Hinohara2014}
Nobuo Hinohara and Jonathan Engel, Phys. Rev. C, {\bf 90}, 031301 (2014).

\bibitem{Pang1969}
Sing~Chin Pang, Nucl. Phys. A, {\bf 128}, 497 -- 526 (1969).

\bibitem{Evans1981}
J.~A. Evans, G.~G. Dussel, E.~E. Maqueda, and R.~P.~J. Perazzo, Nucl. Phys. A,
  {\bf 367}, 77 -- 94 (1981).

\bibitem{Goodman1979}
A.~L. Goodman, Adv. Nucl. Phys., {\bf 11}, 263 (1979).

\bibitem{Guzman2002}
R.~Rodríguez-Guzmán, J.~L. Egido, and L.~M. Robledo, Nucl. Phys. A, {\bf
  709}, 201--235 (2002).

\bibitem{Romero2019}
A.~M. Romero, J.~Dobaczewski, and A.~Pastore, Phys. Lett. B, {\bf 795}, 177 --
  182 (2019).

\bibitem{Kanada2014}
Yoshiko Kanada-En'yo and Fumiharu Kobayashi, Phys. Rev. C, {\bf 90}, 054332
  (2014).

\bibitem{Dobaczewski1984}
J.~Dobaczewski, H.~Flocard, and J.~Treiner, Nucl. Phys. A, {\bf 422}, 103 --
  139 (1984).

\bibitem{STOITSOV200545}
M.~V Stoitsov, Jacek Dobaczewski, Witold Nazarewicz, and Peter Ring, Comput.
  Phys. Commun., {\bf 167}, 45--63 (2005).

\bibitem{Homma1996}
H.~Homma, E.~Bender, M.~Hirsch, K.~Muto, H.~V. Klapdor-Kleingrothaus, and
  T.~Oda, Phys. Rev. C, {\bf 54}, 2972--2985 (1996).

\bibitem{Cohen1965}
S.~Cohen and D.~Kurath, Nucl. Phys., {\bf 73}, 1 -- 24 (1965).

\bibitem{Lisetskiy1999}
A.~F. Lisetskiy, R.~V. Jolos, N.~Pietralla, and P.~von Brentano, Phys. Rev. C,
  {\bf 60}, 064310 (1999).

\bibitem{TILLEY19951}
D.~R. Tilley, H.~R. Weller, C.~M. Cheves, and R.~M. Chasteler, Nucl. Phys. A,
  {\bf 595}, 1--170 (1995).

\end{thebibliography}

\end{document}